\documentclass[fleqn,10pt,a4paper]{wlscirep}
\pdfoutput=1
\usepackage{graphicx}
\usepackage{color}
\usepackage{amssymb}
\usepackage[english]{babel}
\usepackage{amsmath}
\usepackage{pdfpages}

\begin{document}

%\title{Plasmon blockade from metallic nanodimers}
\title{Visible quantum plasmonics from metallic nanodimers}

\renewcommand{\vec}[1]{\mathbf{#1}}
\newcommand{\modal}[1]{\tilde{\vec{#1}}}
\newcommand\addition[1]{#1}

\author[1,+]{F.~Alpeggiani}
\author[2]{S.~D'Agostino}
\author[3]{D.~Sanvitto}
\author[1,*]{D.~Gerace}
\affil[1]{Dipartimento di Fisica, Universit\`a di Pavia, via Bassi 6, 27100 Pavia, Italy}
\affil[2]{Center for Biomolecular Nanotechnologies @ UNILE - Istituto Italiano di Tecnologia, 73010 Arnesano, Italy}
\affil[3]{CNR NANOTEC -- Institute of Nanotechnology, Via Monteroni, 73100 Lecce, Italy}
\affil[+]{Present address: FOM Institute AMOLF, Science Park 104, 1098 XG Amsterdam, The Netherlands.}
\affil[*]{dario.gerace@unipv.it}

\begin{abstract}
We report theoretical evidence that bulk nonlinear materials weakly interacting with highly localized plasmonic modes in ultra-sub-wavelength metallic nanostructures can lead to nonlinear effects at the single plasmon level in the visible range. In particular, the two-plasmon interaction energy in such systems is numerically estimated to be comparable with the typical plasmon linewidths. Localized surface plasmons are thus predicted to exhibit a purely nonclassical behavior, which can be clearly identified by a sub-Poissonian second-order correlation in the signal scattered from the quantized plasmonic field under coherent electromagnetic excitation. We explicitly show that systems sensitive to single-plasmon scattering can be experimentally realized by combining electromagnetic confinement in the interstitial region of gold nanodimers with local infiltration or deposition of ordinary nonlinear materials. We also propose configurations  that could allow to realistically detect such an effect with state-of-the-art technology, overcoming the limitations imposed by the short plasmonic lifetime.
\end{abstract}

\flushbottom
\maketitle

\thispagestyle{empty}

\section*{Introduction}

Strongly confined electromagnetic fields at the nanoscale allow to enhance radiation-matter interaction in the optical or near-infrared domain to such a level that  scattering of single radiation quanta may become important. On the one hand, this is a key target of modern nanophotonics research, e.g., for the realization of all-optical sensing, switching, and routing at the lowest possible control power, up to the level of a single energy quantum \cite{OBrien2009}. On the other hand, it constitutes a potential playground to explore fundamental quantum manybody physics related to strongly correlated states in open systems \cite{Hartmann_review}. Within this context, single-photon nonlinear optics has been predicted in a variety of nanophotonic systems \cite{Tian1992,Imamoglu99,ciuti06prb,chang07np,Gerace2009,Ferretti2012,arka2013prb,rabl2011prl}, and demonstrated through the single-photon blockade effect, i.e., inhibition of the absorption of a single quantum of energy due to the presence of another one, both in atomic \cite{Birnbaum2005} and solid-state \cite{faraon08nphys,Lang2011,Reinhard2012} cavity-quantum-electrodynamics (CQED). In these realizations, a single quantum emitter --- an atom or a quantum dot --- strongly interacts with a single confined mode of a suitably engineered electromagnetic resonator. 
%STE 
%All these systems typically rely on long lifetimes, thus limiting the maximum single-photon emission rates. Here we propose a system where the intrinsically small material nonlinearity \cite{boyd_book} can be enhanced due to electromagnetic field confinement, such that the regime of quantum nonlinearity can be reached even without the use of long-lifetime quantum emitters, potentially providing an unprecedentedly fast source of quantum states of radiation to be eventually employed in quantum information.

Among the different routes to the electromagnetic confinement, plasmonic systems allow for enhancement of radiation-matter interaction to scales well below the diffraction limit, hardly reachable with conventional optical means. In particular, surface plasmons, i.e., mixed radiation-matter excitations that arise at metal-dielectric interfaces due to the coupled oscillation of the electromagnetic field and the electron charge density \cite{Plasmonics_book}, are currently attracting considerable interest to develop quantum plasmonic devices \cite{QPlasmonics_review,Hecht2013}. One of the ultimate goals is to explore the use of surface plasmon excitations to enhance nonlinear interactions with single quantum emitters \cite{Ridolfo2010,Chang2006,Akimov2007}.
From the point of view of classical electromagnetism, optical bistability in metal nanoantennas involving a third-order nonlinear medium has been extensively analyzed \cite{xia}. On the other hand, in the limit of extreme localization of the plasmonic field, a truly quantum regime can be reached, characterized by the nonlinear interaction among single quanta of energy. Only a few pioneering works have proposed to reach such regime of single-plasmon blockade by using non-resonant material nonlinearities. \addition{In particular, early attempts of measuring a plasmonic analog of the Coulomb blockade effect have been reported in Ref.~\cite{Smolyaninov} by detecting quantized steps in the low-power transmission through sub-wavelength metallic pinholes.} More recently, theoretical proposals were focusing on confined surface plasmons in monolayer graphene nanocavities \cite{Manjavacas2012,Gullans2013}, to realize a plasmon blockade effect in the mid-infrared range. \addition{As continuous-wave excitation is generally assumed, the actual observability of such an effect in a realistic experimental set-up may be strongly hindered by the extremely short plasmonic lifetimes, especially when turning to conventional noble metals at optical frequencies. In general, there has been little consideration on the possibility of overcoming these limitations with pulsed excitation techniques.}

\begin{figure}[t]
\centering
\includegraphics{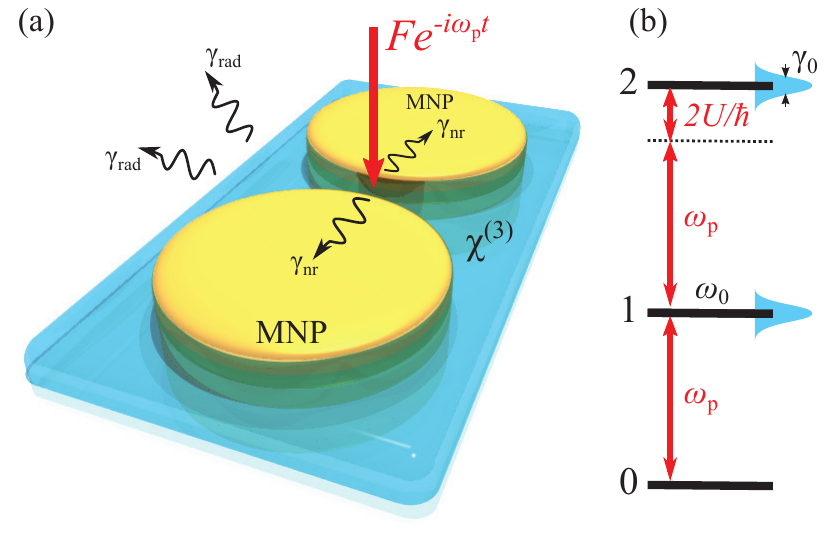}
\caption{(a) Sketch of the system under consideration in this work. A nanodimer made up of two metallic nanoparticles (MNPs) is embedded in a nonlinear medium with a third-order susceptibility $\chi^{(3)}$ and it is coherently pumped at frequency $\omega_p$ from the exterior. The dimer sustains a localized surface plasmon with a decay rate $\gamma_0 = \gamma_{\mathrm{rad}} + \gamma_{\mathrm{nr}}$, due to radiation losses and dissipation. (b) Scheme of the energy levels of the system, highlighting the single-plasmon-blockade mechanism.}\label{fig0}
\end{figure}

%STE
At variance from the conventional CQED systems, which typically rely on long lifetimes, thus limiting the maximum single-photon emission rates, here we propose a system where the intrinsically small material nonlinearity \cite{boyd_book} can be enhanced due to electromagnetic field confinement, so that a regime of quantum nonlinearity can be reached even without the use of long-lifetime quantum emitters. \addition{This potentially provides an extremely fast source of quantum states of radiation with order-of-magnitude improvements over full-dielectric systems, to be eventually employed in quantum information.}
Specifically, in the present work we analyze the quantized surface plasmonic excitations of a metal nanodimer, a nanostructure largely within reach of state-of-the-art technology \addition{in terms of size and gaps between its constitutive elements \cite{Schlather2013,Santhosh2016,Wang2016,Zhu2016}.}
Localized surface plasmons are characterized by a finite linewidth due to both radiation losses and dissipation inside the metal. We show that the single-plasmon blockade can be reached quite straightforwardly under an external coherent excitation, e.g., an external laser input, by infiltrating the interstitial region of the dimer with a sufficiently nonlinear optical material (e.g., molecular dye solutions), as sketched in Fig.~\ref{fig0}(a). \addition{Even if a similar system was proposed in Ref.~\cite{Smolyaninov}, the actual probing of plasmon blockade in terms of quantum correlation measurements, which we theoretically address in the present work, was not considered before.}

% FIG. 1

Although it is easy to compute the response of the metallic nanoparticle with different kinds of external excitations, such as plane waves or oscillating dipoles, establishing a quantum-mechanical theory of the plasmonic excitations is a far more complex task, due to the intrinsically lossy (radiative and non-radiative) character of surface plasmons \cite{spaser,Ridolfo2010,Hummer2013,Gullans2013}. In the following, we show that, by relying on the formalism of quasinormal modes, it is indeed possible to rigorously demonstrate that the system can be modeled with a quantum master equation involving a single bosonic operator, similarly to non-absorbing photonic systems. Within this theoretical picture, the physical basis of the effect can be grasped by looking at the first excitation levels of the system, which are schematically reported in Fig. \ref{fig0}(b). We suppose to coherently pump the nanodimer at a frequency $\omega_p$, in resonance with the bare surface plasmon energy $\hbar\omega_0$. The simultaneous presence of two highly-confined plasmons in the spatial region occupied by the nonlinear material is associated with a large nonlinear interaction, $U$, which produces a shift of the double-excitation energy. The magnitude of $U$ essentially depends on the degree of localization of the electromagnetic field, which can be quantified through the effective volume of the plasmonic mode, $V_{\mathrm{eff}}$. \addition{As we discuss in the following, although the formal expression for $V_{\mathrm{eff}}$ is the same as for dielectric systems, it is essential to ensure the correct normalization of the plasmon eigenfield in agreement with the theory of quasinormal modes.} Moreover, due to their intrinsically lossy character, plasmonic levels present a finite linewidth, $\gamma_0$. When the effective volume is sufficiently small (for the system under consideration, $V_{\mathrm{eff}} \lessapprox 10^{-3}$ $\mu$m$^3$), the magnitude of $U$ could result in a shift of the two-plasmon excitation frequency ($\Delta\omega = 2U/\hbar$, see the scheme in Fig.~\ref{fig0}) that is larger than the natural linewidth of the mode, calculated to be around 100 meV for the typical nanodimers under consideration.  As a result, the system cannot absorb a second plasmon but upon re-emission of the first one, thus effectively becoming a source of single plasmonic excitations.
  
Finally, we theoretically show that single-plasmon blockade could be experimentally measured, despite the fs-scale plasmon lifetime, by using a pulsed excitation source. \addition{Besides opening another route to the experimental study of quantum plasmonic effects, truly meant as the mutual interaction of single plasmon excitations at the nanoscale, these results could enable an ultra-high-rate source of single radiation quanta at visible wavelengths.}

\section*{Results and discussion}

% FIG. 1
\begin{figure}[t]
\centering
\includegraphics[scale=0.85]{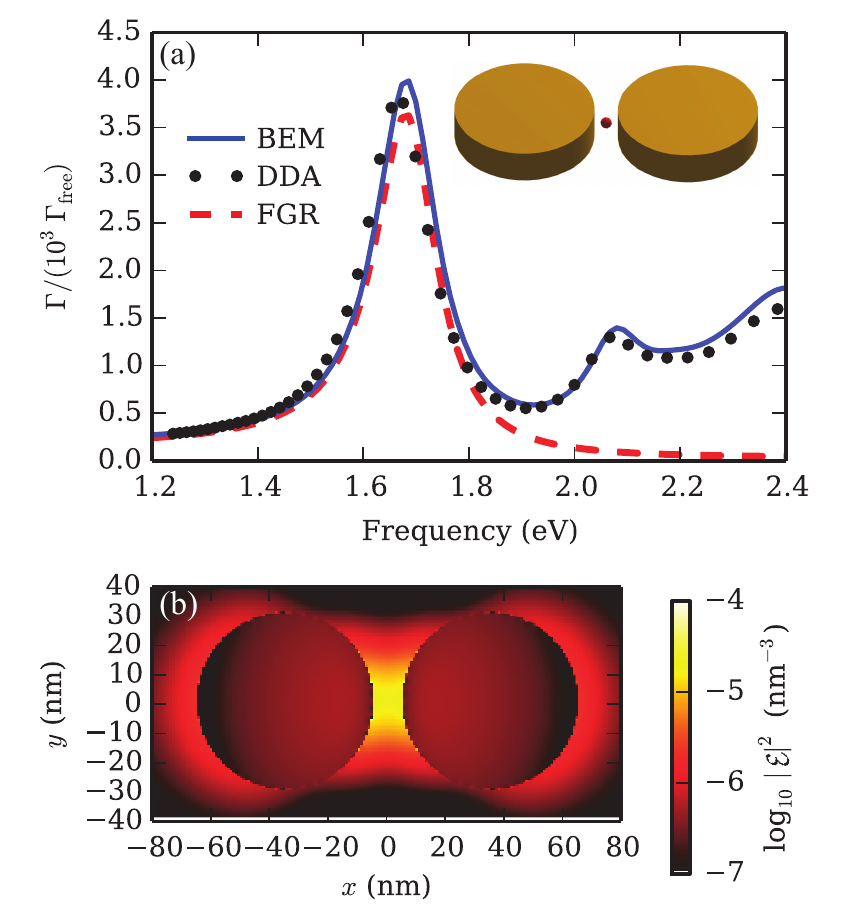}
\caption{(a) Normalized decay rate of a dipolar emitter located at the center of the 10-nm interstitial region between two 60-nm-diameter, 15-nm-high gold nanodisks (see inset), \textit{vs} the emitter energy. The calculation is performed with the BEM, DDA, and Fermi golden rule (FGR) applied to Eq.~\eqref{eq:E}. (b) Cross section of the intensity of the QNM electric field, $|\mathcal{E}|^2$, along the dimer middle plane. The units for the QNM field are chosen in agreement with Eq.~\eqref{eq:E}.}\label{fig1}
\end{figure}

% STE: REMOVE ALL THE SUBSECTIONS TITLES
%\subsection{Localized plasmon modes}
We consider localized surface plasmons of gold nanodimers in the context of the boundary element method (BEM) formalism developed in Ref.~\citenum{bem}, which can be applied to any system of locally homogeneous regions of space separated by abrupt interfaces. In Fig.~\ref{fig1}(a) we plot the semiclassical decay rate of a dipole emitter with momentum $\vec{p}$ located at the center of the 10-nm gap between two gold nanodisks and directed along the inter-particle axis (see inset), showing a clear peak around $\omega = 1.7$ eV due to coupling with a (longitudinal) surface plasmon of the nanoantenna. In this work, we use the experimental dielectric function of evaporated gold \cite{olmon} and we assume a background refractive index $n_{\text{B}} = 1.5$, to account in an averaged way for the surrounding dielectrics. The rate computed with the MNPBEM toolbox (blue solid curve), a publicly available implementation of the BEM \cite{mnpbem}, is compared with a completely independent calculation (dots) within the discrete dipole approximation (DDA) \cite{stefania_dda,adda}, showing very good agreement. Both quantities are normalized to the dipole free-space decay rate $\Gamma_{\mathrm{free}} = \omega^3 n_{\mathrm{B}} p^2 / (3\pi\varepsilon_0\hbar c^3)$.

It is recognized that the definition of localized plasmons in metal nanostructures, and, more generally, of the natural oscillation modes of leaky optical systems, can be made rigorous in the framework of quasinormal modes (QNMs) \cite{hughes12,lalanne1,review}, i.e., the solutions of a non-Hermitian differential equation with complex eigenfrequencies $\tilde{\omega}$. We have computed QNMs with the following procedure. In the BEM formalism, the electric field in each region is decomposed into an incident and a scattered part, $\vec{E}(\vec{r},\omega) = \vec{E}_{\mathrm{inc}}(\vec{r},\omega) + \vec{E}_{\mathrm{sc}}(\vec{r},\omega)$, the latter being identified with the field generated by a (fictitious) charge and current distribution on the enclosing surface. After discretizing the surface into a collection of $N$ representatives points, the charge and current distribution is calculated from the solution of a linear problem of the form
$\Sigma\,\vec{x} = \vec{a}$,
where $\Sigma(\omega)$ is a $N \times N$ matrix and $\vec{a}$ is a vector constructed from the incident field. For brevity, we omit the full equations of the method and we refer to Ref.~\citenum{bem} for the definition of the involved quantities (see Supplementary Information). For our purposes, it suffices to notice that the QNMs of the system are obtained by the condition $\det\Sigma(\tilde{\omega}) = 0$, which corresponds to the \emph{nonlinear eigenvalue problem}
\begin{equation}\label{eq:eigen}
\Sigma(\tilde{\omega})\,\tilde{\vec{x}} = 0
\end{equation}
for the generalized eigenvector $\tilde{\vec{x}}$. More importantly, the quasinormal field (QNF), $\mathcal{E}(\vec{r})$, can be computed as the electric field generated by the surface charge and current distribution obtained from the eigenvector (see Supplementary Information). We have solved Eq.~\eqref{eq:eigen} working within the MNPBEM toolbox and using the two-sided Rayleigh functional iterative algorithm \cite{eigenproblem}. For instance, for the plasmon associated to the peak in Fig.~\ref{fig1}(a), we obtain $\tilde{\omega} \simeq 1.68 - i0.07$ eV, and the QNF intensity plotted in Fig.~\ref{fig1}(b).

At variance from the well-known relation $\int \mathrm{d}\mathbf{r} \, \varepsilon(\vec{r}) |\mathcal{E}(\vec{r})|^2 = 1$, which holds for normal modes in non-dispersive media, QNMs satisfy a much more complex normalization condition \cite{hughes12,lalanne1,review}. Instead of explicitly calculating the normalization integral, we follow an implicit approach to normalize the field, similar to that proposed in Ref.~\citenum{lalanne2}. Indicating with $\vec{r}_0$ the position of the dipole emitter at the center of the dimer gap, we normalize the QNM by imposing the following relation \cite{lalanne1,lalanne2} for $\omega \to \tilde{\omega}$:
\begin{equation}\label{eq:norm}
\vec{E}_{\mathrm{sc}}(\vec{r},\omega) \approx
- \omega \frac{\vec{p}\cdot\mathcal{E}(\vec{r}_0)}{2\varepsilon_0
(\omega - \tilde{\omega})}\mathcal{E}(\vec{r})
\end{equation}
(the factor $2\varepsilon_0$ is for dimensional convenience). We further improved the procedure by factoring out the singular term and imposing the relation only to the residues, with great advantages in terms of numerical accuracy and computational efficiency (see Supplementary Information).

%STE: THIS SECTION MUST BE REDUCED OR SIMPLIFIED
%\subsection{Theoretical methods}
The usual starting point for the quantization of plasmonic systems is the system-bath approach \cite{Hummer2013}, where the electric field is expanded on a continuum of frequency-dependent bosonic operators. Following Ref.~\citenum{Hummer2013} and introducing the collective modes $\hat{p}(\omega)$, Eq.~\eqref{eq:norm} implies that for $\omega \approx \omega_0$ and $\gamma_0 r / c \ll 1$ the field operator can be approximated as
\begin{equation}
\hat{\vec{E}}(\vec{r},t) \simeq i\sqrt{\frac{\hbar \omega_0}{2\varepsilon_0}}
\mathcal{E}(\vec{r})
\int \mathrm{d}\omega\,
\frac{g(\omega)}{\sqrt{2\pi}}\,
\hat{p}(\omega)e^{-i\omega t} + \mathrm{H. c.},
\end{equation}
with $g^2(\omega) = \gamma_0 / [(\omega - \omega_0)^2 + \gamma_0^2/4]$ and $\mathcal{E}(\vec{r})$ the QNF introduced previously (see Supplementary Information). This model corresponds to a structured bosonic bath with a Lorentzian spectral density, whose central frequency $\omega_0 = \mathrm{Re}(\tilde{\omega})$ and linewidth $\gamma_0 = -2\mathrm{Im}(\tilde{\omega})$ are directly related to the QNM eigenfrequency. According to a well-known theoretical result \cite{pseudomode,Hummer2013}, the dynamics of such a model is equivalent to that of a single bosonic mode, $\hat{p}$, which can thus be interpreted as the plasmon destruction operator, coupled to a flat reservoir with dissipation rate $\gamma_0$. 
Thus, the electric field operator can be equivalently written
\begin{equation}\label{eq:E}
\hat{\vec{E}}(\vec{r},t) = i\sqrt{\frac{\hbar \omega_0}{2\varepsilon_0}}
\left[\hat{p}e^{-i\omega_0 t}\mathcal{E}(\vec{r}) - \hat{p}^{\dagger}e^{i\omega_0 t}\mathcal{E}^*(\vec{r})\right],
\end{equation}
with the system dynamics being described by a density-matrix master equation in the Markov approximation: 
\begin{equation}\label{eq:master}
i \hbar \dot{\rho} = [\hat{H}, \rho] + \frac{i \, \hbar \gamma_0}{2} [2\hat{p}\rho\hat{p}^{\dagger} - \hat{p}^{\dagger}\hat{p}\rho - \rho\hat{p}^{\dagger}\hat{p}].
\end{equation}
This represents a commonly employed approach in the recent field of quantum plasmonics \addition{\cite{spaser,Ridolfo2010,Hummer2013,Gullans2013}}. 

To check its validity, in Fig.~\ref{fig1} we also show (red dashed curve) the emitter decay rate calculated in the weak coupling regime by applying the Fermi golden rule to Eq.~\eqref{eq:E}, i.e., $\Gamma(\omega) = \omega_0 \gamma_0 |\vec{p}\cdot\mathcal{E}(\vec{r}_0)|^2 / [2 \varepsilon_0 \hbar ((\omega - \omega_0)^2 + \gamma_0^2/4)]$ and normalized to $\Gamma_{\mathrm{free}}(\omega)$.
The very good agreement with the semi-classical result near the resonance frequency confirms the validity of our model and justifies the single-mode approximation of Eq.~\eqref{eq:E}.

We assume the plasmonic dimer to be embedded in a nonlinear medium, e.g., with a strong third-order nonlinear susceptibility $\chi^{(3)}$, and we write the Hamiltonian in the form
\begin{equation}\label{eq:hamiltonian0}
\hat{H}_0=\hbar\omega_0 \hat{p}^{\dagger}\hat{p}  + U \hat{p}^{\dagger}\hat{p}^{\dagger}\hat{p}\hat{p} \, 
\end{equation}
where the first term accounts for the energy of the plasmon mode, including the contribution of metal dispersion (as implied by the QNM normalization condition).  The second term is a four-operator product deriving from the third-order susceptibility of the nonlinear medium around the dimer and accounting for two-plasmon scattering. In a sense, this term is the analog of Hubbard interaction in electron systems. If we assume the $\chi^{(3)}$ to be nonzero only in a region outside the nanodimers and dispersionless, we can use the same expression for the nonlinear interaction energy already introduced for dielectric systems~\cite{Ferretti2012}: 
\begin{equation}\label{eq:nonlinearity}
U= \frac{(\hbar\omega_0)^2 } {\varepsilon_0}  \int \mathrm{d}\mathbf{r} \,\, \chi^{(3)}(\vec{r}) |\mathcal{E}(\vec{r})|^4.
\end{equation}
\addition{The master equation approach would also allow to include nonlinear loss terms, such as the nonlinear absorption introduced by an imaginary part of the $\chi^{(3)}$ response, which we neglect due to the small pumping rates considered here (see, e.g., Ref.~\cite{Ferretti2012}). }
Moreover, we are considering a scalar nonlinear response (i.e., no tensorial coupling between the different $\mathcal{E}$ components), since \addition{we are mainly interested in a proof-of-principle theoretical demonstration of the quantum nonlinear behavior in a standard experimental setup as a function of the possible values of the $\chi^{(3)}$ elements.}
In this respect, we can simplify the nonlinear coupling energy to $U = (\hbar\omega_0)^2 \chi^{(3)} V_{\mathrm{eff}}^{-1} / \varepsilon_0 $, where the effective volume of the confined plasmon is defined as
\begin{equation} \label{eq:vol_eff}
V_{\mathrm{eff}}^{-1} = \int_{\mathrm{\Omega}} \mathrm{d}\mathbf{r} \,\, |\mathcal{E}(\vec{r})|^4,
\end{equation}
the integration being limited to the domain $\Omega$ represented by the bounded physical volume filled with the nonlinear medium. $V_{\mathrm{eff}}$ essentially reflects the degree of localization of the QNF in region $\Omega$. \\
\addition{As a final approximation, we are neglecting any nonlocal effects of the $\chi^{(3)}$ response, which are known to arise for gap sizes certainly smaller than 1 nm \cite{Ciraci,Mortensen}. We also stress that recent experimental investigations at the single or few molecules level within the sub-nm gap size between a metallic microsphere and a planar surface have been thoroughly explained by purely local theoretical modeling \cite{Baumberg}. }
 
% FIG 2
\begin{figure}[t]
 \begin{center}
\includegraphics[scale=0.9]{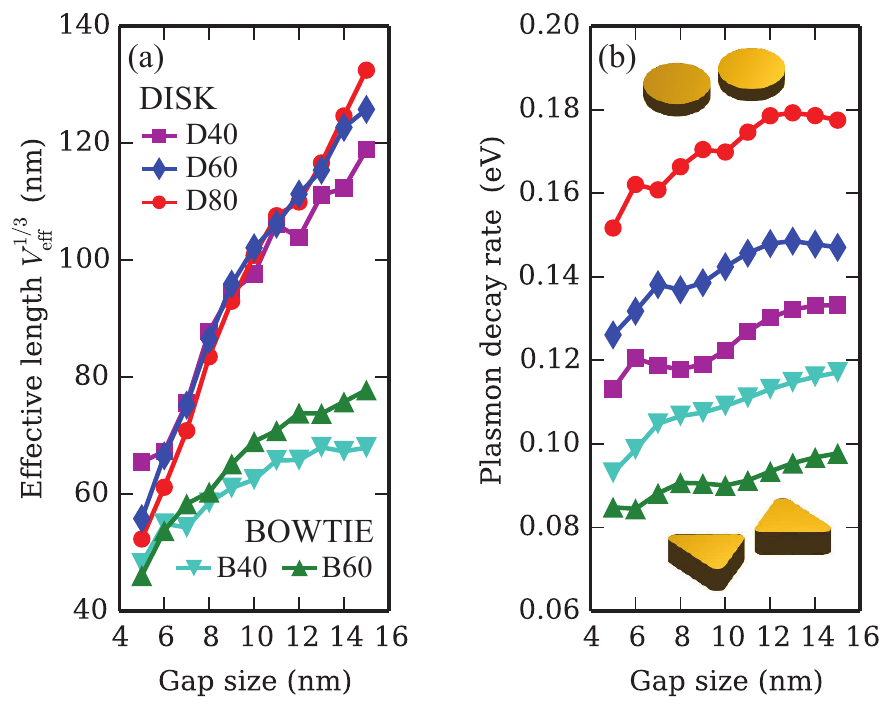}
   \caption{(a) Effective length, $V_{\mathrm{eff}}^{1/3}$, and (b) decay rate, $\hbar\gamma_0$, for the lowest-frequency longitudinal plasmon in dimers of different gold nanoparticles (see insets), versus the interstitial gap size. The curves refer to the case of disks with diameter of 40, 60, and 80 nm and bowtie dimers obtained from equilater triangles with edge of 40 and 60 nm (curvature radius $=$ 5 nm). The height of all particles is 15 nm.}\label{fig2}
\end{center}
\end{figure}

By solving the eigenproblem \eqref{eq:eigen} within the MNPBEM toolbox, we have characterized the lowest-frequency longitudinal plasmon modes of two kinds of nanodimers: nanodisk dimers and bowtie dimers made of two mirroring triangular nanoparticles, for various sizes and inter-particle distances. The results for the effective length $V_{\mathrm{eff}}^{1/3}$ and the linewidth $\hbar\gamma_0$ are summarized in Fig.~\ref{fig2}(a) and (b), respectively. The effective length depends mostly on the gap size, reflecting the plasmonic field enhancement in the interstitial region, whereas the linewidth is mainly affected by the particle size (due to radiative decay) and plasmon frequency (due to metal dissipation). Single plasmon blockade is based on a delicate interplay between $V_{\mathrm{eff}}$ and $\gamma_0$, thus both quantities must be taken into account when choosing the best geometry. It is already evident, however, that bowtie dimers are favored with respect to nanodisks. \addition{In any case, we notice that even the smallest gap size considered in our simulations, i.e. 5 nm, is much larger than what is currently achieved \cite{Wang2016,Zhu2016}.}
The integral in Eq.~\eqref{eq:vol_eff} has been performed in the region external to the metal and bounded by a box spanning the same height of the dimer along $z$ and extending 10 nm beyond the particle extremities in the $xy$ plane. This choice is compatible, for instance, with a film of nonlinear medium coated on top of a substrate. In general, we observed little variation of the values of the effective length when varying the bounding box, as long as the interstitial region between the metal nanostructures is entirely included in the integration volume. We also notice that the structures under consideration are well described within the regime of classical electromagnetic theory, and do not require to include quantum effects due to the microscopic nature of the metallic surfaces \cite{Zuloaga2009,Marinica2012}.

% FIG 3
\begin{figure}[h]
 \begin{center}
  \includegraphics[scale=1]{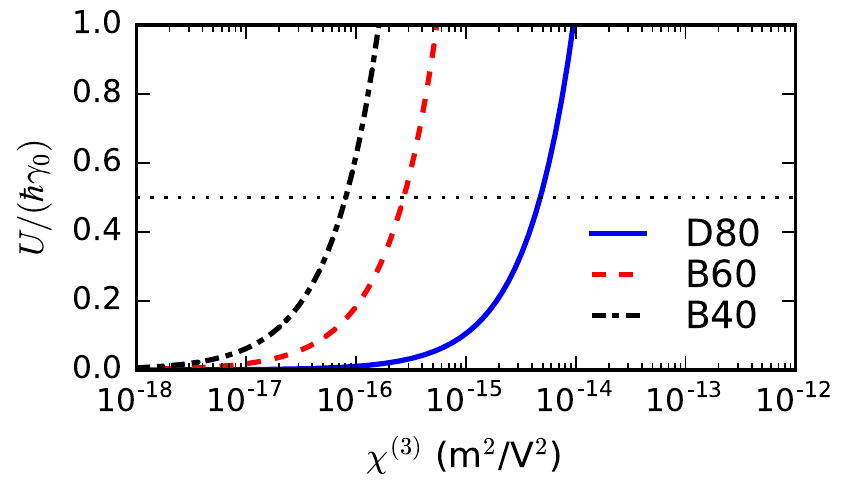}
   \caption{\addition{Ratio between the nonlinear coupling energy and the plasmon linewidth, $U / (\hbar\gamma_0$), as a function of the third-order nonlinear susceptibility of the infiltrated nonlinear medium, for some structures of Fig. \ref{fig2}. Gap sizes are: 15 nm (D80), 10 nm (B60), and 5 nm (B40). The horizontal dotted line corresponds to the value $U / (\hbar\gamma_0) = \tfrac{1}{2}$.}
}  \label{ratio}
\end{center}
\end{figure}

%\subsection{Single-plasmon blockade}.
\addition{Following our previous considerations, the key quantity for quantifying the occurrence of blockade effects in the system is the ratio between the double-excitation energy shift and the plasmon linewidth, or, equivalently, the ratio $U / (\hbar\gamma_0)$. In Fig.~\ref{ratio} we plot this ratio as a function of the nonlinear susceptibility of the infiltrated medium} for a few cases of potential interest, selected from the results presented in Fig.~\ref{fig2}. Three possible nanoplasmonic dimers are considered, either with disk or bowtie geometry. \addition{We expect that plasmon blockade starts to play a significant role in the quantum dynamics of the system when the energy shift becomes comparable with the linewidth, i.e., when $U / (\hbar\gamma_0) > \tfrac{1}{2}$. As it can be seen from Fig.~\ref{ratio}, this condition corresponds to values of the nonlinear susceptibility $\chi^{(3)} \gtrsim 10^{-16}\,\mathrm{m}^2/\mathrm{V}^2$.}

\begin{figure}[t]
 \begin{center}
  \includegraphics[width=0.43\textwidth]{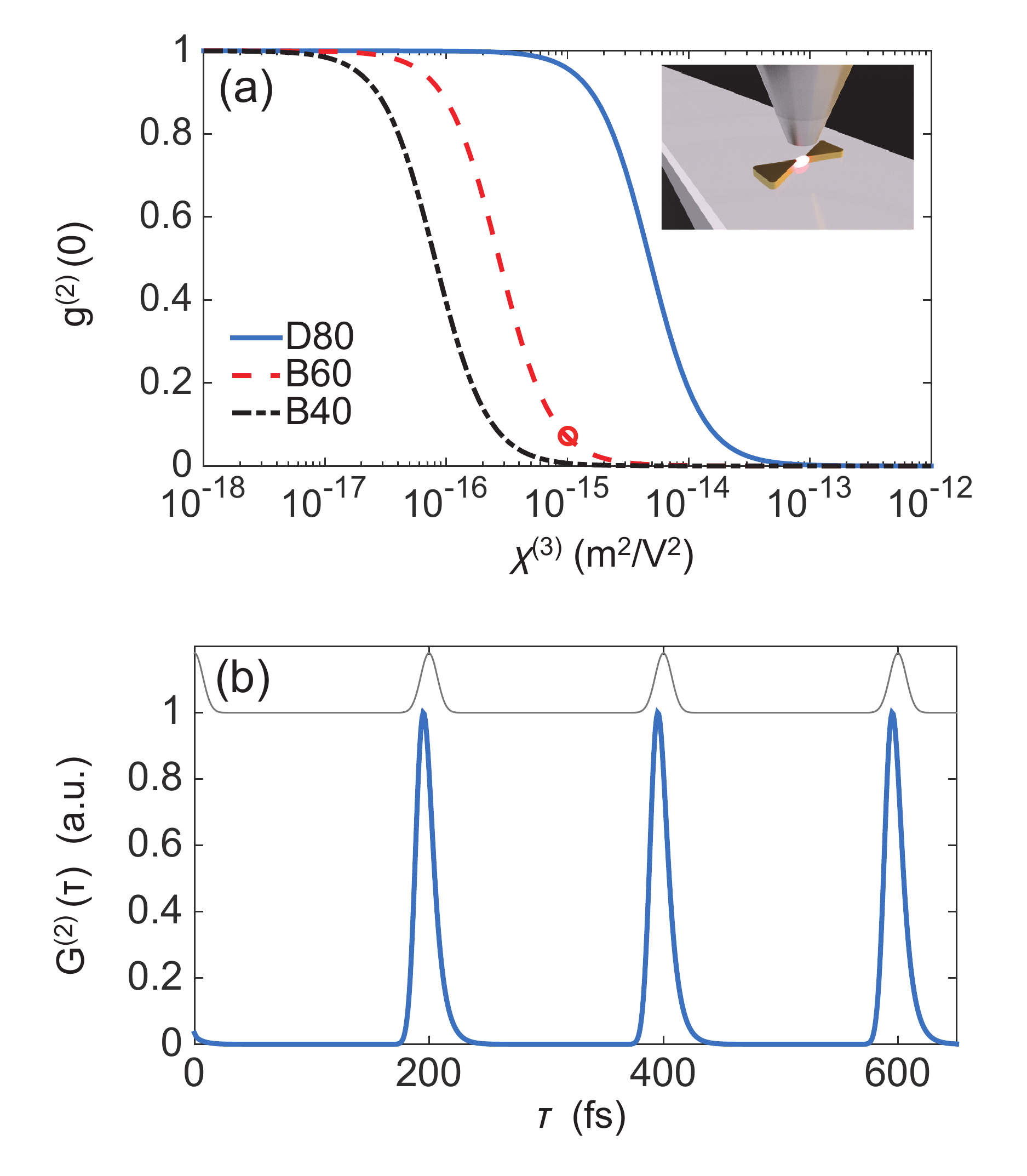}
   \caption{(a) Second-order correlation function at zero time delay as a function of the third-order nonlinear susceptibility magnitude, \addition{for the same structures as Fig.~\ref{ratio}}. A schematic representation of the proposed experiment is given in the inset. (b) Second-order correlation under pulsed excitation for the B60 structure and $\chi^{(3)}=10^{-15}$ m$^2$/V$^2$ (corresponding to the circled result in the previous panel). The pulse sequence is represented in the top of the figure.
}  \label{fig3}
\end{center}
\end{figure}

\addition{Such predictions are confirmed by detailed numerical solutions of the quantum master equation \eqref{eq:master} based on the modeling above.} In order to take into account the realistic implementation of a quantum nanoplasmonic device, we assume an experimental configuration allowing for a coherent driving of the localized plasmon, which can be described by the Hamiltonian
\begin{equation}\label{eq:hamiltonian1}
\hat{H}=\hat{H}_{0} + F e^{-i\omega_p t} \hat{p}^{\dagger}+F^{\ast} e^{i\omega_p t}\hat{p}  \,,
\end{equation}
where $F(t)/\hbar$ represents the effective rate of plasmon excitation at the fixed external frequency $\omega_p$ (e.g., imposed by a pumping laser). As a figure of merit for quantum nonlinear behavior, we focus on the degree of antibunching in the second-order plasmon correlations \cite{loudon_book}, $G^{(2)}(t,t') = \langle \hat{p}^{\dag} (t) \hat{p}^{\dag} (t') \hat{p}(t') \hat{p}(t) \rangle$. In Fig.~\ref{fig3}(a) we plot the normalized function at zero time delay ($\tau=t' - t=0$), defined as $g^{(2)}(0) = G^{(2)}(0)/(\langle \hat{p}^{\dag} \hat{p} \rangle)^2$, under continuous wave excitation ($F/\hbar=0.01 \gamma_0$) and as a function of the nonlinear susceptibility of the infiltrated nonlinear material, which is supposed to have the same refractive index $n_\mathrm{B} = 1.5$ of the background (see Supplementary Information). In all cases, the condition $g^{(2)}(0) < 0.5$ can be considered as a threshold for single-plasmon blockade, in analogy to photon counting statistics in CQED experiments \cite{Birnbaum2005,faraon08nphys,Reinhard2012}. Ideally, when $g^{(2)}(0) \to 0$ the probability of detecting two photons at the output of the device is negligible soon after having detected one. This corresponds to the single-plasmon-blockade regime, which we have schematically shown in Fig.~\ref{fig0}(b). We assume that the external driving field is in resonance with the bare plasmonic mode. Indeed, this corresponds to the condition of maximum plasmon antibunching \cite{Ferretti2012}. In general, increasing the detuning between the pump and the plasmonic resonance is detrimental for the observation of the blockade effect, especially for the case of positive detuning ($\omega_p > \omega_0$), which could result in the direct excitation of the two-plasmon state blue-shifted by the nonlinear interaction.

% FIG 3

%STE
%Under such conditions, the system behaves as a \textit{single-plasmon source}, as schematically indicated in the inset of Fig.~\ref{fig3}(a): when the nonlinear shift introduced by the presence of two plasmons is larger than the linewidth, the system can absorb a second plasmon only upon re-emission of the first one, thus effectively becoming a source of single plasmonic excitations. 
\addition{In agreement with the more qualitative analysis of Fig.~\ref{ratio}}, values of the order of $\chi^{(3)}\sim 10^{-16}$ m$^2$/V$^2$ are predicted from the results of Fig.~\ref{fig3}(a) to be sufficient to reach the single-plasmon blockade regime in the structure with the tightest electromagnetic field confinement, i.e. the bowtie structure with a 5-nm gap between the triangle tips. Such nonlinear values can be achieved, e.g., in glasses doped with metallic nanoparticles \cite{martinez2010,ganeev_apB}. In such cases, care must be taken in the choice of metallic doping, since the plasmon excitations of the nanoparticles embedded in the glass matrix might interact with the target surface plasmon excitation of the nanodimer, thus making the whole analysis more complicated.
In any case, stronger nonlinearities are necessary for structures with inter-particle gaps larger than 10 nm. On the other hand, such high values of the third-order nonlinearity might be within reach of available materials. For instance, values on the order of $\chi^{(3)}\sim 10^{-15} -10^{-13}$ m$^2$/V$^2$ have been measured in organic dye molecules in the visible range \cite{Dye_nonlinear_1,Dye_nonlinear_2}. \addition{Therefore, the use of organic molecules appears a promising route for the experimental observation of plasmon blockade.} In order to avoid complications due to strong interaction between surface plasmons and optical excitations of the nonlinear medium, \addition{as well as avoiding nonlocality of the nonlinear response}, it is important to select a frequency range characterized by a significant third-order nonlinearity together with an essentially flat first-order response.

These results suggest that the single-plasmon blockade regime might be within reach. However, experimentally detecting this effect might still be challenging due to the extremely short plasmon lifetime. For a decay rate $\hbar \gamma_0  \sim 100$ meV, one expects $\tau_p = 1/\gamma_0 \sim 6.6$ fs: this is beyond any possible resolution of single-counting detection. However, such an experiment could still be performed under pulsed excitation, where sufficiently short pulses would overcome the issue of short lifetime under cw excitation. In Fig.~\ref{fig3}(b) we show that this is indeed possible: a train of gaussian pulses with 10 fs duration is sent on the structure B60 assuming $\chi^{(3)}\sim 10^{-15}$ m$^2$/V$^2$ [see Fig.~\ref{fig3}(a)]. The suppression of the zero-time delay peak in the unnormalized function $G^{(2)}(\tau=t'-t)$, calculated with the master equation after a two-time super-operator evolution \cite{ciuti06prb,arka2013prb} (see Supplementary Information for details), is the signature of single-plasmon blockade in this system (in this simulation, $t=12$ fs). In particular, this device demonstrates a single-plasmon source on demand, since each pulse triggers the re-emission of a single-plasmon from the system before a second one can be excited. Given the short plasmon lifetimes, an ultra-high emission rate can be achieved in principle (about 5 THz in this case), \addition{beyond state-of-art demonstrations based on plasmon enhancement of single emitters spontaneous emission rate \cite{Hoang}}. As such, this system can be turned into an ultra-fast single-plasmon source conditioned to the availability on a seemingly fast laser source. For instance, ultra-high repetition rate femtosecond lasers have been shown up to 200 GHz range \cite{Peccianti}. \addition{We notice that the usefulness of such an ultra-fast source could be currently hindered by the lack of sufficiently fast single-photon detectors, although a ps-time resolution for photon pair correlation measurements has been shown by use of a streak camera-based technique in the visible/near-infrared range \cite{Assman}. On the other hand, such a fast system response might be employed, e.g., in engineering novel quantum devices, such as single-photon switches or ultra-fast single photon detectors, which we leave for further exploration.} \\
A possible experimental configuration is schematically described in the inset of Fig.~\ref{fig3}(a): a waveguide below the nanostructured dimer excites the localized plasmon mode, and scattered radiation is collected from a near-field tip (e.g., a scanning near-field optical microscope probe), where the output is sent to a Hanbury Brown-Twiss set-up for correlation measurements. Coincidences counts from the outgoing pulses would reveal the plasmon blockade regime, similarly to experiments performed with CQED systems \cite{faraon08nphys,Reinhard2012}.

In summary, we have theoretically shown the potential interest of metallic nanodimers interacting with ordinary nonlinear media to achieve the quantum plasmonic regime, as defined by the true interaction between two plasmonic quanta confined in the same spatial region. The main contributions given in this work are twofold: on the one hand, we have provided an effective method to estimate the single-plasmonic confinement in the hot-spot of the dimer nanostructures. On the other hand, we have shown how the single-plasmon blockade can be evidenced by time-resolved second-order correlation measurements under pulsed excitation, thus overcoming the intrinsically short plasmon lifetimes, which we believe might stimulate further experimental research in quantum plasmonics, turning short lifetime into a resource.

\section*{Acknowledgements}
The authors acknowledge Darrick Chang, Milena De Giorgi, Antonio Fernandez-Dominguez, Martijn Wubs, for very useful discussions and highly appreciated remarks and suggestions, and F. Della Sala for making computational facilities available. DG is indebted to B. Hecht for initially motivating this investigation. The work was partially supported by the ERC grant ``Polaflow''.

%%%%%%% supplementary material %%%%%%%%%%%%%%%
\newpage


\section*{Supplementary material}

# Supporting Information:
# Visible quantum plasmonics from metallic nanodimers

F. Alpeggiani,[1, *] S. D'Agostino,[2] D. Sanvitto,[3] and D. Gerace[1, †]

[1]*Dipartimento di Fisica, Università di Pavia, via Bassi 6, 27100 Pavia, Italy*
[2]*Center for Biomolecular Nanotechnologies @ UNILE - Istituto Italiano di Tecnologia, 73010 Arnesano, Italy*
[3]*CNR NANOTEC - Institute of Nanotechnology, Via Monteroni, 73100 Lecce, Italy*

## COMPUTING QUASINORMAL MODES

We hereby give the detailed procedure employed to compute the characteristics of localized surface plasmons in metallic dimers. Surface plasmons are identified with quasi-normal modes (QNMs), i.e., the solutions of a non-Hermitian electromagnetic problem, which are calculated in the context of the boundary element method (BEM) described in Ref. 1. All calculations are performed within the MNPBEM toolbox [2, 3], which constitutes a freely available implementation of the BEM.

According to the BEM formalism of Ref. 1, space is subdivided in several regions $V_j$ ($j = 1, 2, \dots$) with locally homogeneous dielectric functions (in our case, the inner region of the dimer and the external space). The electric field in each region is decomposed into an incident and a scattered part, the latter being identified with the field generated by (fictitious) charge and current distributions $\sigma_j(\mathbf{r})$ and $\mathbf{h}_j(\mathbf{r})$ on the enclosing surface, as follows:

$$\mathbf{E}_{\rm sc}(\mathbf{r}, \omega) = i\frac{\omega}{c} \int_{\partial \Omega_j} G_j(\mathbf{r} - \mathbf{s}) \mathbf{h}_j(\mathbf{s}) \, {\rm d}\mathbf{s} - \nabla \int_{\partial \Omega_j} G_j(\mathbf{r} - \mathbf{s}) \sigma_j(\mathbf{s}) \, {\rm d}\mathbf{s}. \tag{1}$$

$G_j(\mathbf{r}) = \exp(ik_j|\mathbf{r}|)/|\mathbf{r}|$ is the scalar Green function for region $j$. Distributions $\sigma_j(\mathbf{r})$ and $\mathbf{h}_j(\mathbf{r})$ are the unknowns of the problem, and they must be computed from boundary conditions at the separating interfaces.

After the surface is discretized into a set of $N$ simpler polygons (triangles and quadrilaterals), surface charge and current distributions become $N$-dimensional vectors. As illustrated in the text, the QNMs of the system are obtained by solving the nonlinear eigenvalue problem [4]

$$\Sigma(\tilde{\omega}) \tilde{\mathbf{x}} = 0 \tag{2}$$

for the (complex) eigenvalue $\tilde{\omega}$ and the right eigenvector $\tilde{\mathbf{x}}$. The matrix $\Sigma$ is defined after Eq. (22) of Ref. 1. Once the eigenvector $\tilde{\mathbf{x}}$ is known, the corresponding charge and current distributions are obtained from the following expressions:

$$\begin{aligned} \sigma_j &= G_j^{-1} \tilde{\mathbf{x}}, \\ \mathbf{h}_j &= i\frac{\omega}{c} \hat{\mathbf{n}} \, G_j^{-1} \Delta^{-1} (L_1 - L_2) \tilde{\mathbf{x}}, \end{aligned} \tag{3}$$

where $j = 1, 2$ indicates the inner and outer regions of the particle, respectively, $\hat{\mathbf{n}}$ is the normal vector to each surface element, and we still refer to Ref. 1 for the definition of the matrices $G_j$, $L_j$, and $\Delta$. The quasinormal field $\mathcal{E}(\mathbf{r})$ is calculated (up to a normalization constant that will be fixed in the following) by replacing the charge and current distributions of Eqs. (3) into Eq. (1).

We have solved the nonlinear eigenproblem (2) with the two-sided Rayleigh functional iterative algorithm [4], which allows to compute the eigenvalue and the right and left eigenvectors, indicated as $\tilde{\mathbf{x}}$ and $\tilde{\mathbf{y}}$, respectively. Starting from an initial triplet $(\tilde{\mathbf{x}}_0, \tilde{\omega}_0, \tilde{\mathbf{y}}_0)$, at each iteration $k$ the values of the three quantities is updated with respect to the previous estimate by solving the equations:

$$\Sigma(\tilde{\omega}_k) \, \mathbf{x}_{k+1} = \Sigma'(\tilde{\omega}_k) \, \mathbf{x}_k; \tag{4}$$

$$[\Sigma(\tilde{\omega}_k)]^* \, \mathbf{y}_{k+1} = [\Sigma'(\tilde{\omega}_k)]^* \, \mathbf{y}_k; \tag{5}$$

$$\tilde{\omega}_{k+1} = \tilde{\omega}_k - \frac{\mathbf{y}_{k+1}^* \, \Sigma(\tilde{\omega}_k) \, \mathbf{x}_{k+1}}{\mathbf{y}_{k+1}^* \, \Sigma'(\tilde{\omega}_k) \, \mathbf{x}_{k+1}}. \tag{6}$$

The notation $\Sigma'$ indicates the derivative $d\Sigma/d\omega$, which can be approximated by finite differences by storing the values of the previous iteration, whereas $\Sigma^*$ indicates the conjugate transpose. In order to speed up convergence, the initial guess $\tilde{\mathbf{x}}_0$ can be extracted from the solution of the electromagnetic problem with a suitable external excitation (e.g.,

a dipole emitter) near resonance. In addition, the vectors $\tilde{\mathbf{x}}$ and $\tilde{\mathbf{y}}$ can be renormalized at each iteration to reduce overflow errors. As the eigenfrequency $\tilde{\omega}$ is typically complex, we had to modify the MNPBEM toolbox to extend the refractive index of the particle to the complex domain, by means of a first-order analytical continuation of the form

$$n_{\text{met}}(\omega' + i\omega'') \approx n_{\exp}(\omega') + i\omega'' \frac{dn_{\exp}(\omega')}{d\omega'}, \tag{7}$$

where $n_{\exp}$ is a cubic spline interpolation of experimental data and the derivative of the resulting piecewise polynomial is calculated analytically. As stated in the text, we employed the tabulation of the experimental dielectric function of evaporated gold published in Ref. 5.

In order to ensure the proper normalization of the QNM, we follow an implicit approach involving an additional dipole source, similar to that presented in Ref. 6. We add a point dipole source with momentum $\mathbf{p} = p\hat{\mathbf{x}}$ at the position $\mathbf{r}_0$ in the center of the dimer gap (see the inset in Fig. 2 of the main text) and we calculate the field scattered back by the particle to the dipole for $\omega \to \tilde{\omega}$. The normalization condition, reported by Eq. (2) of the main text, reads as follows:

$$\mathbf{E}_{\text{sc}}(\mathbf{r}, \omega) \approx -\omega \frac{\mathbf{p} \cdot \mathcal{E}(\mathbf{r}_0)}{2\varepsilon_0(\omega - \tilde{\omega})} \mathcal{E}(\mathbf{r}). \tag{8}$$

Notice that both terms are proportional to the momentum $\mathbf{p}$. The dependence in the left-hand term is implicit. The condition can be simplified by replacing the matrix $\Sigma^{-1}$ in the calculation of the scattered field with the polar approximation for $\omega \approx \tilde{\omega}$ [4]

$$\Sigma^{-1}(\omega) \approx \frac{1}{\omega - \tilde{\omega}} \frac{\tilde{\mathbf{x}}\tilde{\mathbf{y}}^*}{\tilde{\mathbf{y}}^*[d\Sigma/d\omega]\tilde{\mathbf{x}}}, \tag{9}$$

where $\tilde{\mathbf{x}}$ and $\tilde{\mathbf{y}}$ are the right and left eigenvectors, respectively. This allows to factor out the singular term from both sides of Eq. (8) and to recast the normalization condition in the form

$$\mathbf{p} \cdot \mathcal{E}(\mathbf{r}_0) = -\frac{2\varepsilon_0}{\tilde{\omega}} \frac{1}{\tilde{\mathbf{y}}^*[d\Sigma/d\omega]\tilde{\mathbf{x}}} \tilde{\mathbf{y}}^* \left\{ D^e - \Sigma_1 L_1 \phi^e \right.$$
$$\left. + i\frac{\omega}{c}\hat{\mathbf{n}} \cdot [(L_1 - L_2)\Delta^{-1}(\alpha + ik\hat{\mathbf{n}}L_1\phi^e) + (L_2\Delta^{-1}\Sigma_1 - L_1\Delta^{-1}\Sigma_2)\mathbf{A}^e] \right\}. \tag{10}$$

The term in curly brackets is the same as that in the right-hand side of Eq. (22) in Ref. 1 (where the definition of all included quantities can be found), and it is a source term constructed from the field emitted by the dipole.

Equation (8) does not require any further calculations than those already performed in order to solve the nonlinear eigenproblem, greatly improving the computational efficiency. We employed surface meshes with up to about 6000 elements. Typically, a few iterations of the algorithm are enough to converge to the solution of the eigenproblem. The computational time for the calculation of a single QNM (including normalization) is of the order of a few minutes with a quad-core 3.1 GHz personal computer.

The MNPBEM toolbox has been employed also to calculate the semiclassical decay rate of the dipole emitter of Fig. 2(a) of the main text. The rate is computed through the formula [7]:

$$\Gamma(\omega) = \Gamma_{\text{free}}(\omega) + \frac{2}{\hbar} \text{Im} \left[ \mathbf{p} \cdot \mathbf{E}_{\text{sc}}(\mathbf{r}_0) \right], \tag{11}$$

where $\mathbf{E}_{\text{sc}}(\mathbf{r}_0)$ is the field scattered back by the nanodimer at the dipole position.

## DISCRETE DIPOLE APPROXIMATION

The modification of the dipole emitter total decay rate induced by the presence of the dimer has also been quantitatively assessed in the framework of the discrete dipole approximation (DDA) full-wave simulation method, which describes the scatterer as an array of polarizable points organized on a regular cubic grid [8]. This method yields solutions for the electromagnetic field in response to an incident electric field in the frequency domain, including retardation effects. The polarization of each element internal to the scatterer is seen as the result of the interaction with the local electromagnetic field produced by all the other dipoles plus the external field of the emitter. By simply fixing the dipole position in space, the changes induced by the dimer antenna on the local field experienced by the dipole can be rigorously calculated. The validity of the method been largely tested in recent works [9].

The simulations for the present work have been performed by using the parallel code ADDA in its final version, which implements dipolar sources [10]. The inter-dipole distance is fixed to 0.25 nm - a value that is small enough to achieve numerical convergence - and the gold refractive index is taken from Ref. 5, similarly to the BEM calculations described before.

## QUANTIZATION PROCEDURE FOR QUASINORMAL MODES

In the context of the system-bath approach for the quantization of the electromagnetic field in the presence of dispersive and dissipative media, the electric field operator is expanded on a continuum of "true" normal modes including the effect of the dispersive media, in the following form [11, 12]:

$$\hat{\mathbf{E}}(\mathbf{r}) = i\sqrt{\frac{\hbar}{\pi\varepsilon_0}} \int d\omega \frac{\omega^2}{c^2} \int d\mathbf{r}' \sqrt{\mathrm{Im}\,\varepsilon(\mathbf{r}',\omega)} \overset{\leftrightarrow}{G}(\mathbf{r},\mathbf{r}',\omega)\hat{\mathbf{f}}(\mathbf{r}',\omega) + \mathrm{h.c.}, \tag{12}$$

where $\hat{\mathbf{f}}(\mathbf{r},\omega)$ are (vector) destruction operators satisfying the bosonic commutation relation $[\hat{f}_j(\mathbf{r},\omega), \hat{f}_k^\dagger(\mathbf{r}',\omega')] = \delta(\mathbf{r}-\mathbf{r}')\delta(\omega-\omega')\delta_{jk}$ ($j,k=1,2,3$), and "h.c." indicates the Hermitian conjugate operator. The quantity $\overset{\leftrightarrow}{G}(\mathbf{r},\mathbf{r}',\omega)$ is the dyadic Green function of the (classical) electric field, satisfying $\nabla\times\nabla\times\overset{\leftrightarrow}{G}(\mathbf{r},\mathbf{r}',\omega) - \omega^2\varepsilon(\omega)\overset{\leftrightarrow}{G}(\mathbf{r},\mathbf{r}',\omega)/c^2 = \overset{\leftrightarrow}{I}\delta(\mathbf{r}-\mathbf{r}')$. Each tensor component $G_{jk}$ is defined as the $j$-component of the electric field at point $\mathbf{r}$ calculated in the presence of the dispersive media for a point-like current source at point $\mathbf{r}'$ and of the form: $\mathbf{J}(\mathbf{r}) = -i\delta(\mathbf{r}-\mathbf{r}')\hat{\mathbf{x}}_k/(\omega\mu_0)$. The source corresponds to an oscillating dipole with momentum $\mathbf{p} = \hat{\mathbf{x}}_k/(\omega^2\mu_0)$; as a consequence, we can use the result in Eq. (8) to obtain the expansion of the dyadic Green function in proximity to a QNM:

$$G_{jk}(\mathbf{r},\mathbf{r}',\omega) = -\frac{\omega}{2(\omega-\tilde{\omega})}\frac{c^2}{\omega^2}\mathcal{E}_j(\mathbf{r})\mathcal{E}_k(\mathbf{r}') + G_{\mathrm{other}}(\mathbf{r},\mathbf{r}',\omega). \tag{13}$$

The term $G_{\mathrm{other}}(\mathbf{r},\mathbf{r}',\omega)$ includes the free-space contribution and those of the other QNMs of the system. Since, for $\omega \simeq \mathrm{Re}(\tilde{\omega})$ the dominant term is the pole corresponding to the QNM, we will assume $G_{\mathrm{other}}(\mathbf{r},\mathbf{r}',\omega) \simeq 0$. This assumption is further confirmed by the comparison of the decay rates in Fig. 2 of the main text.

Along the lines of the approach presented in Appendix A.2 of Ref. 12, we introduce the collective modes $\hat{p}(\omega)$ according to the following relation:

$$\mathbf{g}(\mathbf{r},\omega)\hat{p}(\omega) = \sqrt{\frac{\hbar}{\pi\varepsilon_0}}\frac{\omega^2}{c^2}\int d\mathbf{r}' \sqrt{\mathrm{Im}\,\varepsilon(\mathbf{r}',\omega)} \overset{\leftrightarrow}{G}(\mathbf{r},\mathbf{r}',\omega)\hat{\mathbf{f}}(\mathbf{r}',\omega). \tag{14}$$

The modes form a continuum depending only on frequency and satisfy the bosonic commutation relation $[\hat{p}(\omega),\hat{p}^\dagger(\omega')] = \delta(\omega-\omega')$. Moreover, we assume the vectors $\mathbf{g}(\mathbf{r},\omega)$ to be real. These assumptions will be justified at the end of the procedure. Following the calculations in Ref. 12, by replacing Eq. (14) into the commutation relation, we obtain:

$$g_j(\mathbf{r},\omega)g_k(\mathbf{r}',\omega) = \frac{\hbar}{\pi\varepsilon_0}\frac{\omega^2}{c^2}\mathrm{Im}G_{jk}(\mathbf{r},\mathbf{r}',\omega). \tag{15}$$

Then, by further replacing the expansion of Eq. (13) into the latter expression, we arrive at

$$g_j(\mathbf{r},\omega)g_k(\mathbf{r}',\omega) = -\frac{\hbar\omega}{2\pi\varepsilon_0}\mathrm{Im}\left[\frac{\mathcal{E}_j(\mathbf{r})\mathcal{E}_k(\mathbf{r}')}{\omega-\tilde{\omega}}\right] \simeq \frac{1}{2\pi}\frac{\hbar\omega_0}{2\varepsilon_0}\left[\frac{\gamma_0}{(\omega-\omega_0)^2+\gamma_0^2/4}\right]\mathcal{E}_j(\mathbf{r})\mathcal{E}_k(\mathbf{r}') \tag{16}$$

($\tilde{\omega} = \omega_0 - i\gamma_0/2$). As a result, indicating as $g^2(\omega)$ the quantity in square brackets in the last term, we can write

$$\mathbf{g}(\mathbf{r},\omega) = \sqrt{\frac{\hbar\omega_0}{2\varepsilon_0}}\frac{g(\omega)}{\sqrt{2\pi}}\mathcal{E}(\mathbf{r}), \tag{17}$$

which, together with Eqs. (12) and (14), gives Eq. (3) in the main text. In Eq. (16) we suppose that the QNM field $\mathcal{E}(\mathbf{r})$ can be taken approximately as real, as we verified to be the case for all the systems under consideration in this work ($\mathrm{Re}\mathcal{E} \gg \mathrm{Im}\mathcal{E}$ near the gap region). In other systems, this approximation might not hold. This latter situation corresponds to the case of a non-Lorentzian density of states discussed in Ref. 13, which leads to a pathological master equation not in the Lindblad form considered here.

## CALCULATING THE SECOND-ORDER CORRELATIONS

For a given quantized field described by destruction (creation) operators $\hat{A}$ ($\hat{A}^\dagger$), the figure of merit quantifying the single-photon sensitivity is the time-ordered second-order autocorrelation function, defined as [14]

$$G_i^{(2)}(t,t') = \langle \hat{A}^\dagger(t)\hat{A}^\dagger(t')\hat{A}(t')\hat{A}(t)\rangle, \quad (18)$$

where $t' - t = \tau > 0$, which is normalized as

$$g^{(2)}(t,t') = \frac{G^{(2)}(t,t')}{\langle \hat{A}^\dagger(t)\hat{A}(t)\rangle\langle \hat{A}^\dagger(t')\hat{A}(t')\rangle}. \quad (19)$$

In the manuscript, we have provided results for the numerical calculations of the latter quantities as follows. The master equation in Eq. (5) of the main text is recast in the form

$$\frac{d\hat{\rho}}{dt} = \frac{i}{\hbar}[\hat{\rho}, \hat{H}_{\rm rot}] + \frac{\gamma}{2}\left[2\hat{A}\hat{\rho}\hat{A}^\dagger - \hat{A}^\dagger\hat{A}\hat{\rho} - \hat{\rho}\hat{A}^\dagger\hat{A}\right], \quad (20)$$

where the Hamiltonian is rotated with respect to the driving laser frequency and it is represented as

$$\hat{H}_{\rm rot} = \hbar\Delta\,\hat{A}^\dagger\hat{A} + U\,(\hat{A}^\dagger)^2\hat{A}^2 + F(t)\hat{A}^\dagger + F^*(t)\hat{A}, \quad (21)$$

with $\Delta = \omega_0 - \omega_p$. The Hilbert space is truncated to 11 plasmons, largely sufficient for convergence under the weak pump amplitudes assumed in this work ($F_0/\hbar = 0.01\gamma_0$).

Under continuous wave (cw) excitation, $F(t) = F_0$, the steady state zero-time delay second order correlation can be easily calculated as

$$g^{(2)}(0) = \langle(\hat{A}^\dagger)^2\hat{A}^2\rangle/\langle\hat{A}^\dagger\hat{A}\rangle^2 = {\rm Tr}\{(\hat{A}^\dagger)^2\hat{A}^2\hat{\rho}_{ss}\}/n^2, \quad (22)$$

where $n = {\rm Tr}\{\hat{A}^\dagger\hat{A}\hat{\rho}_{ss}\}$, and $\hat{\rho}_{ss}$ is the steady state solution corresponding to $d\hat{\rho}/dt = 0$.

Under pulsed excitation, e.g. for a train of Gaussian pulses described by $F(t) = F_0\exp\{-(t - nT_0)^2/\Delta T^2\}$ (with $n = 0, \pm 1, ...$), where $T_0$ and $\Delta T$ are the pulse separation and width, respectively, Eq. (18) is evaluated numerically through a superoperator evolution [15–17]

$$G^{(2)}(t,t') = {\rm Tr}\{\hat{A}\mathcal{U}_{t,t'}[\hat{A}\hat{\rho}(t)\hat{A}^\dagger]\hat{A}^\dagger\}, \quad (23)$$

where $\mathcal{U}_{t,t'}$ indicates the evolution from $t$ to $t'$ with Eq. 20 when assuming the starting time operator as $\hat{A}\hat{\rho}(t)\hat{A}^\dagger$.

---



\end{document}